\documentclass[12pt,preprint,referee]{aastex}

\usepackage[usenames]{color}

\usepackage{graphics}
\usepackage{graphicx}
\usepackage{epstopdf}
\usepackage{url}
\usepackage[stable]{footmisc}
\usepackage{lscape}

\newcommand{\be}{\begin{equation}}
\newcommand{\ee}{\end{equation}}
\newcommand{\ba}{\begin{eqnarray}}
\newcommand{\ea}{\end{eqnarray}}
\newcommand{\bc}{\begin{center}}
\newcommand{\ec}{\end{center}}

\newcommand{\uu}{4U 1036--56 }
\newcommand{\is}{IBIS/ISGRI }
\newcommand{\je}{JEM--X }

\begin{document}

\title{\textsc{\emph{INTEGRAL} and \emph{Swift} observations of the Be X-ray binary 4U~1036-56 (RX J1037.5-5647) \\
and its possible relation with $\gamma$-ray transients}}
\author{Jian Li\altaffilmark{1,2}, Diego F. Torres\altaffilmark{2,3}, Shu Zhang\altaffilmark{1}, Alessandro Papitto\altaffilmark{2}, \\
Yupeng Chen\altaffilmark{1}, Jian-Min Wang\altaffilmark{1,4,5}}

\altaffiltext{1}{Key Laboratory for Particle Astrophysics, Institute of High Energy Physics, Chinese Academy of Sciences, 19B Yuquan Road, Beijing 100049, China }
\altaffiltext{2}{Institut de Ci\`encies de l'Espai (IEEC-CSIC),
             Campus UAB,  Torre C5, 2a planta,
              08193 Barcelona, Spain}
\altaffiltext{3}{Instituci\'o Catalana de Recerca i Estudis Avan\c{c}ats (ICREA).}
\altaffiltext{4}{National Astronomical Observatories of China, Chinese Academy of Sciences, 20A Datun Road, Beijing 100020, China}
\altaffiltext{5}{Theoretical Physics Center for Science Facilities (TPCSF), Chinese Academy of Sciences, Beijing 100049, China}

\begin{abstract}
We present timing, spectral, and long-term temporal analysis of the high mass X-ray binary (HMXB)
\uu using \emph{INTEGRAL} and \emph{Swift} observations. We show that it is a weak hard X-ray source
spending a major fraction of the time in quiescence, and only occasionally characterized by X-ray outbursts. The outburst activity we report here lasts several days, with a dynamic range spanned by the luminosity in quiescence and in outburst as high as $\sim30$. We report the detection of pulse period at 854.75$\pm$4.39 s during an outburst, which is consistent with previous measurements. Finally, we analyze the possibility of 4U 1036--56's association with the unidentified transient $\gamma$-ray sources AGL J1037--5708 $\&$ GRO J1036--55, as prompted by its positional correlation.

\end{abstract}

\keywords{X-rays: binaries, X-rays: individual (4U 1036--56)}

\section{Introduction}

Various classes of $\gamma$-ray sources, from Galactic objects like supernova remnants, pulsar wind nebulae, and binaries,  to starburst galaxies and distant blazars (see, e.g., Hinton $\&$ Hofmann 2009), have been detected up to the TeV band.
Among them, the class of $\gamma$-ray emitting binaries focuses great attention.
They are X-ray binaries hosting O/B companions, which have $\gamma$-ray emission up to High-Energy (HE, E$\textgreater$100 MeV) and/or Very High-Energy (VHE, E$\textgreater$100 GeV) regimes, modulated on the orbital period.
Only a handful of such binaries are known (e.g., LS 5039, LSI +61 303, PSR B1259-63, HESS J0632+057, 1FGL J1018.6-5856, or Cyg X-3), although a larger population is expected (see, e.g., Ackermann et al. 2012).

In recent years, a number of unidentified, transient $\gamma$-ray
 sources were discovered in the Galactic plane, especially by {\it Astrorivelatore Gamma a Immagini Leggero (AGILE)}. Many of them have also been suggested to have a possible
 binary nature (see, e.g., Sguera 2011a and references therein, Casares
 et al. 2012, see Torres et al. 2001 for variability studies in the
 EGRET era). These are shown in Table 1. All of them show transient
 behavior in $\gamma$-rays, and most of them are observed having fast
 and strong X-ray activity. No blazar-like candidate counterparts are
 found within their positional error uncertainties. Instead, all six
 candidates have been suggested to have a possible high mass X-ray
 binary (HMXB) counterpart, four of which are confirmed supergiant fast X-ray
 transients (SFXT) or candidate SFXTs. Two of the HXMB candidates are already identified
 to host a slowly rotating pulsar. The probability of finding e.g., a
 supergiant HMXB inside the {\it AGILE} error circle by chance, given the
 number of supergiant HMXBs detected by IBIS on the {\it International Gamma-Ray Astrophysics Laboratory (INTEGRAL)} within the Galactic plane
 is $\sim 1$\%, i.e. $\sim$ 0.5 chance coincidences are expected
 (Sguera et al. 2011a).

Two $\gamma$-ray transients (AGL J1037-5708 and GRO J1036-55) are located in the same region of the sky, and with the caveat of the large positional uncertainties involved, they could perhaps be associated (see Figure 1).
AGL J1037-5708 is an
unidentified, transient MeV source discovered by {\it AGILE} near the Galactic plane (Bulgarelli et al. 2010). {\it AGILE} detected intense $\gamma$-ray emission above
 100 MeV from AGL J1037-5708 between 2010-11-27 21:18 UT to 2010-11-30 14:08 UT. A maximum likelihood analysis yields a detection  larger than 5 $\sigma$ for energies larger than 100 MeV with a flux above 300 $\times10^{-8}$ photons  cm$^{-2}$ s$^{-1}$.

\begin{landscape}
\begin{table*}
\centering
\scriptsize
\caption{Unidentified transient $\gamma$-ray sources in the Galactic plane with possible binary nature. }
\begin{tabular}{llllllll}
\hline
\hline
\\
Name              & Possible X-ray       & Possible optical or           & Distance (kpc)  &Refs.   & Duration of flares (outburst)&Dynamic range & Refs.   \\
Refs.             & counterpart          &  infrared counterpart                 & Orb. period (days) &     & &  \\
                  &                      &                             & Orb. eccentricity  &     & &  \\
                  &                      &                             & Pulse period (s)  &     &  & \\

\hline
\\
AGL J1734$-$3310  &IGR J17354$-$3255   & 2MASS                      & $\sim 8.5$    	& 2 & 0.5--65 hours @ 18--60 keV & $\sim$ 200& 1, 4 \\
1, 2, 3, 4         &  transient HMXB     & J17352760                  & 8.4474$\pm$0.0017 & 1 & 1 day @ $E >100$  MeV & ?&3 \\
                 & candidate SFXT      & $-$3255544                 &  ?	            &  &   \\
                 &  		           & 			                &  ?		        &  &   \\

\hline
\\
AGL J2022$+$3622 &IGR J20188$+$3647    &  ?              & ?       &	    & 60 mins @ 20--40 keV & $\sim$ 90&5 \\
5, 6           &  transient          &                    & ?    &     & 1 day @ E$>100$  MeV & ? &6\\
                 & candidate SFXT       &                   &  ?	 &             &   \\
                 &  		           & 			        &  ?	&	          &   \\

\hline
\\
ERG J1122$-$5946 & IGR J11215$-$5952 &  HD 306414   &  $\sim 6.2$& 10 & 15 mins--2 hours (15 days) @ 1--10 keV & $\textgreater$ 1000&7, 8 \\
5, 7, 8, 9,        &  transient HMXB pulsar       &   (B1 Ia)   & $\sim 164.6$ &7 &  variable @ $E>100$  MeV & ?&12\\
10, 11, 12                  &confirmed SFXT      &                &  ?	           &   &   \\
                 &  		           & 			    &  186.78$\pm$ 0.3		&9          &   \\

\hline
\\

AGL J2241$+$4454 & ?                       &  MWC 656             &  2.6$\pm$  1.0    & 14    && ? \\
 13, 14, 15        &  binary system          &(B3 IVne$+$sh)        & 60.37$\pm$ 0.04 &14 &1 day @ E$>100$  MeV  &?&13\\
                 &  with compact star      &                      & 0.4  $\pm$ 0.1    & 15      &   \\
                 &  		               & 			          &   ?               &   	 &   \\

\hline
\\

AGL J1037$-$5708   &  4U 1036$-$56             &  LS 1698  &  $\sim 5$  & 18  & (4--10 days) @ 18--60 keV & $\sim 30$ &21\\
GRO J1036$-$55     &  persistent HMXB pulsar   &(B0 IIIVe) & 847? or 645?  &19  &  3 days @ $E >100$  MeV & ?&16\\
16, 17, 18, 19         &                           &           & ?             &   & $\sim$ 12 days @ 3--10  MeV & ?&17 \\
20, 21, 22        &  		                   & 	       & 853.4$\pm$0.2 & 20  &   \\

\hline
\\

3EG J1837--0423  & AX J1841.0-0536     & 2MASS      &  6.9$\pm$1.7  &  24         &   a few days (a few hours) @ 0.2--10 keV & $\sim 1600$ & 27 \\
 23, 24, 25    &transient HMXB       & J18410043  &  ?                &       &   3.5 days @ $E >100$  MeV & ?&23 \\
26, 27, 28      &confirmed SFXT      &  -0535465  &  ?                &   &       \\
		      &                     & (B1 Ib)    &  ?    & 25 &      &       \\

\hline

\end{tabular}
\label{expo-hard}
\tablecomments{
  (1) Sguera et al. 2011a
  (2) Tomsick et al. 2009
  (3) Bulgarelli et al. 2009
  (4) Sguera et al. 2011b
  (5) Sguera 2009b
  (6) Chen et al. 2007
  (7) Romano et al. 2009  
  (8) Romano et al. 2007  
  (9) Swank et al. 2007  
  (10) Masetti et al. 2006  
  (11) Negueruela et al. 2005  
  (12) Casandjian J-M. $\&$ Grenier 2008  
  (13) Lucarelli et al. 2010  
  (14) Williams et al. 2010  
  (15) Casares et al. 2012  
  (16) Bulgarelli et al. 2010  
  (17) Zhang $\&$ Collmar 2007  
  (18) Motch et al. 1997  
  (19) Sarty et al. 2011  
  (20) La Palombara et al. 2009  
  (21) this paper
  (22) Krimm et al. 2012  
  (23) Tavani et al. 1997 
  (24) Sguera et al. 2009a 
  (25) Bozzo et al. 2011 
  (26) Nespoli et al. 2008 
  (27) Romano et al. 2011 
  (28) Sguera V. et al. 2006 
}
\end{table*}
\end{landscape}

GRO J1036--55 was discovered by the {\it Imaging Compton Telescope (COMPTEL)} on {\it Compton Gamma-Ray Observatory (CGRO)} at a significance level of 5.6 $\sigma$, reaching a flux level of 350 mCrab in the 3-10 MeV band (Zhang $\&$ Collmar 2007). During the COMPTEL's lifetime of 9 years, GRO J1036--55 is only visible during that flare, from 1996--10--03 to 1996--10--15 (MJD 50359--MJD 50361). An analysis of simultaneous {\it Energetic Gamma Ray Experiment Telescope (EGRET)} data at energies larger than 100 MeV did not yield any evidence for the source. The energy spectrum indicated a spectral maximum at $\sim 4$ MeV.

The HXMB \uu is the only X-ray source located in the region of these transients.
\uu first appeared in the 4th UHURU catalogue (Forman et al. 1978), and it was detected by the {\it Seventh Orbiting Solar Observatory (OSO-7)} as 1M 1022-554 (Markert et al. 1979). The first X-ray outburst from \uu was observed by Ariel V in 1974, from November 11th to 19th (MJD 42362--MJD 42370), see Warwick et al. (1981).
The maximum flux of the outburst was 2.4 $\times 10^{-10}$ erg cm$^{-2}$ s$^{-1}$, 2.4 times the Uhuru average flux in the same energy range (2--10 keV).
During the {\it Roentgen Satellite (ROSAT)} survey, \uu was about 10 times dimmer than it was in 1970--1976 (Motch et al. 1997). \uu was later observed by {\it Rossi X-ray Timing Explorer (RXTE)} (Reig $\&$ Roche, 1999), {\it X-ray Multi-Mirror Mission (XMM-Newton)} (La Palombara et al. 2009), {\it Swift} (see below for details), and appeared in the third and fourth IBIS catalogues (Bird et al. 2007, 2010). The flux
measured by\textit{ XMM-Newton} is also one order of magnitude lower
than the previous average value, suggesting \uu is characterized
by significant variability.\footnote{One of the \textit{ROSAT} observations yielded an even lower luminosity but may possibly be inaccurate because it is derived from a count rate measurement in a different energy range, then converted into 2-10 keV energy band (La Palombara et al. 2009).}
\textit{Chandra, Suzaku,} and \textit{Monitor of All-sky X-ray Image (MAXI)} have not observed \uu yet.
Thus, across a time interval
of about 35 years, \uu was detected at a luminosity of (1--3) $\times 10^{35}$ erg s$^{-1}$
between 2 and 10 keV in several instances, with some excursions to lower values (La Palombara et al. 2009).
The optical counterpart of \uu is identified with LS 1698, a B0 III--Ve star at $\sim$5 kpc (Motch et al. 1997).
Reig \& Roche (1999) performed detailed timing and spectral analysis with \textit{RXTE} observations,
and discovered a pulsation with period P = 860 $\pm$ 2 s. Based on this behavior, Reig $\&$ Roche (1999) proposed that \uu is a persistent,
low-luminosity binary pulsar.
With \textit{XMM-Newton} observations, La Palombara et al. (2009) found a period of 853.4$\pm$0.2, indicating an average pulsar spin-up $\dot P \sim -2 \times 10^{-8}$ s s$^{-1}$ in the last decade.

  Except for \uu and MWC 656 (see Table 1), the other four possible X-ray counterparts for $\gamma$-ray transients are studied in detail in the
  literature (see Sguera et al. 2009a,b, 2011a).
  But opposite to MWC 656, for which few X-ray observations are available
  and no X-ray counterpart is identified, there is plenty of data in X-rays for 4U 1036--56 that has yet to be analyzed. In this work, we report on timing, spectrum, and long-term temporal analysis of {\it INTEGRAL} and {\it Swift} data on 4U 1036--56, examining the possibility of its association with the unidentified transient $\gamma$-ray sources AGL J1037--5708 $\&$ GRO J1036--55.

\section{Observations and data analysis}

{\it INTEGRAL} (Winkler et al. 2003) is a $\gamma$-ray mission covering the energy the range 15 keV--10 MeV.
Observations of {\it INTEGRAL} are
carried out in individual Science Window (ScW), which have a typical time
duration of about 2000 s. For the {\it INTEGRAL} analysis in this paper, we use all public \is and \je data for which \uu has offset angle less than 9$^o$ and 5$^o$, respectively.
Our data set comprised about 1607 ScWs for \is and 856 ScWs for JEM--X. The data covers revolutions 36--867, from 2003-01-29 to 2009-11-20 (MJD 52668--55155), adding up to a total exposure time of 4.42 Ms for \is and 1.02 Ms for \je (0.91 Ms from JEM-X 1 and 0.1 Ms from JEM-X 2). The data reduction is performed using the
standard ISDC offline scientific analysis software version 9.0. \is images for each ScW are generated in the energy band of 18--60 keV. The counts rate at the position of the source are extracted from all individual images to produce the long--term lightcurve on the ScW time--scale.
When \uu is found to be in outburst, the spectra are produced following the standard steps as stated in the IBIS Analysis User Manual, running the pipeline from the raw data to SPE and LCR level. In the quiescent period, the spectrum is obtained using the mosaic images, as is appropriate for spectral analysis of faint sources.
The spectrum and lightcurve of \je during the outburst period is produced following to the standard steps as stated in JEM-X Analysis User Manual, using OSA 9.0.\footnote{See \url{http://www.isdc.unige.ch/integral/analysis} for more information.}

{\it Swift} (Gehrels et al. 2004) is a $\gamma$-ray burst explorer. It carries three co-aligned detectors: the Burst Alert Telescope (BAT, Barthelmy et al. 2005), the X-Ray Telescope (XRT, Burrows et al. 2005), and the
Ultraviolet/Optical Telescope (UVOT, Roming et al. 2005).
To compare with {\it INTEGRAL} results, the 65--months
{\it Swift}/BAT Snapshot Survey lightcurve of \uu is also inspected.\footnote{See \url{http://swift.gsfc.nasa.gov/docs/swift/results/bs58mon/SWIFT_J1037.6-5649} for more information on {\it Swift}/BAT survey lightcurve of 4U 1036-56.} The 14--195 keV lightcurve covers from 2004-12-16 (MJD 53355) to 2010-05-31 (MJD 55347). It is produced with individual snapshot images from $\sim$5-minute observations, corrected for off-axis effects. This lightcurve is rebinned to a 1-day timescale to improve the signal to noise ratio. The total exposure time of \uu in this lightcurve is 19.4 Ms.

We have also used the data collected by the XRT instrument on board the {\it Swift} satellite (when the source was in outburst, see Krimm et al. 2012).
 \uu was first detected by the BAT in the 15-50 keV band on 2012 Feb. 3, its flux peaked on 2012 Feb. 6 (MJD 55963), remaining
 detectable through Feb. 13.
A 2982-seconds, Photon Counting mode XRT observation was performed on 2012 Feb. 17, for which we present timing and spectral analysis.
We use archival level 2 XRT data in our analysis.
The selection of event grades is 0-12, for Photon Counting data (see Burrows et al. 2005).
To correct for the pile-up effect we estimate the size of the Point Spread Function (PSF) core affected. By comparing the observed and nominal PSF (Romano et al. 2006; Vaughan et al. 2006), a radius of 4 pixels are determined and all the data within this radius
from \uu are excluded. Source events are accumulated within an annulus (inner radius of 4 pixels and outer radius of 30 pixels, 1 pixel $\sim$ 2.36 arcsec)\footnote{See \url{http://www.swift.ac.uk/analysis/xrt/pileup.php} for more information}.
Background events are accumulated within a circular, source-free
region with a radius of 60 pixels.
For timing analysis, the BARYCORR task is used to perform barycentric corrections to the photon arrival times. We extract lightcurves with a time resolution of 2.5 seconds. The XRTLCCORR task is used to account for the pile-up correction in the background-subtracted light curves.
For our spectral analysis, we extract events in the same regions as those adopted for the lightcurve creation.
Exposure maps are generated with the task XRTEXPOMAP. Ancillary response files are
generated with the task XRTMKARF, to account for different extraction regions,
vignetting, and PSF corrections.
In order to search for a periodic signal in the {\it Swift} lightcurve, we
used the Lomb--Scargle periodogram method (Lomb 1976; Scargle
1982). Power spectra are generated for the lightcurve
using the PERIOD subroutine (Press $\&$ Rybicki 1989). The 99.99\%
white noise significance level is estimated
using Monte Carlo simulations (see e.g. Kong, Charles $\&$
Kuulkers 1998).
The 99\% red noise significance level is
estimated using the REDFIT subroutine, which can provide the red noise
spectrum via fitting a first-order autoregressive process to the
time-series (Schulz $\&$ Mudelsee 2002; Farrell et al. 2009).\footnote{See \url{ftp://ftp.ncdc.noaa.gov/pub/data/paleo/softlib/redfit} for more information.}
All of the spectral analysis is performed using XSPEC version
12.6.0; uncertainties are given at the 1$\sigma$ confidence level for
one single parameter of interest.

\section{Results}

Combining all the ISGRI data, \uu is detected by \is  with a
significance level of 11.2 $\sigma$ and an average intensity of 0.180$\pm$0.016 counts s$^{-1}$ in the 18--60 keV band. Figure 1 (left panel) shows the \is mosaic image of the \uu sky region. In Figure 1, the position of the transient {\it AGILE} source AGL J1037--5708 ($>$ 100 MeV) (Bulgarelli et al. 2010) is also plotted, noted with its 95$\%$ position uncertainty as a white circle. Also, the transient source GRO J1036--55 detected by COMPTEL in the 3-10 MeV band (Zhang $\&$ Collmar 2007) is shown in the sky region, noted with its 1, 2, and 3 $\sigma$ positional uncertainty by the green lines.
\uu is detected by JEM-X at 3-20 keV at a significance of 4 $\sigma$ (combining all data from JEM--X1 and JEM--X2), with an average intensity of $0.88\pm0.22 \times 10^{-4}$ counts/cm$^{2}$/s.
The \je mosaic image of this sky area is shown in Figure 1 (right panel).

We investigated the \is long-term lightcurve of \uu on the ScW timescale in the 18-60 keV band, see Figure 2. During most of the time, \uu is not significantly detected by \is at the ScW level, its significance being below 2 $\sigma$ (dotted blue line in Figure 2, upper panel). However, 12 ScWs have a significance larger than 4$\sigma$ and all of them are located in a period of {about} 5 days, between MJD 54142 (2007 Feb. 11) and MJD 54147 (2007 Feb. 16), which is highlighted, labeled, and zoomed in Figure 2. Hereafter, we refer to this period as the {\it INTEGRAL} outburst.

The recent outburst observed by {\it Swift} (Krim et al. 2012) is also marked and labeled in Figure 2. There is no simultaneous \is observation by {\it INTEGRAL}.  The time of MeV flare from the transient $\gamma$-ray source AGL J1037-5708 is noted in Figure 2 too. This period also lacks simultaneous observations by {\it INTEGRAL}.

Finally, we have also inspected the {\it Swift}/BAT survey lightcurve. In 19.4 Ms, \uu is detected with 8.2 $\sigma$ only (derived from the daily binned lightcurve). The average flux is
(3.03$\pm0.37)\times10^{-5}$ counts s$^{-1}$, in the 14-195 keV band. {The significances of the daily lightcurve are Gaussian-distributed, with a mean value of 0.172 and a deviation (1 $\sigma$) of 1.17. Under this distribution and considering the 1775 points forming the daily lightcurve, there should be less than one observation with a single trial significance larger than 4$\sigma$. However, several such are discovered and they located at MJD 54144, MJD 54148, right at the {\it INTEGRAL} outburst period, and at MJD 54445, MJD 54499, MJD 54501, and MJD 54890. Since {\it Swift}/BAT is not as sensitive as IBIS/ISGRI, the lightcurve  is not appropriate for further analysis.

\begin{figure*}[t]
\centering
\includegraphics[angle=0, scale=0.43] {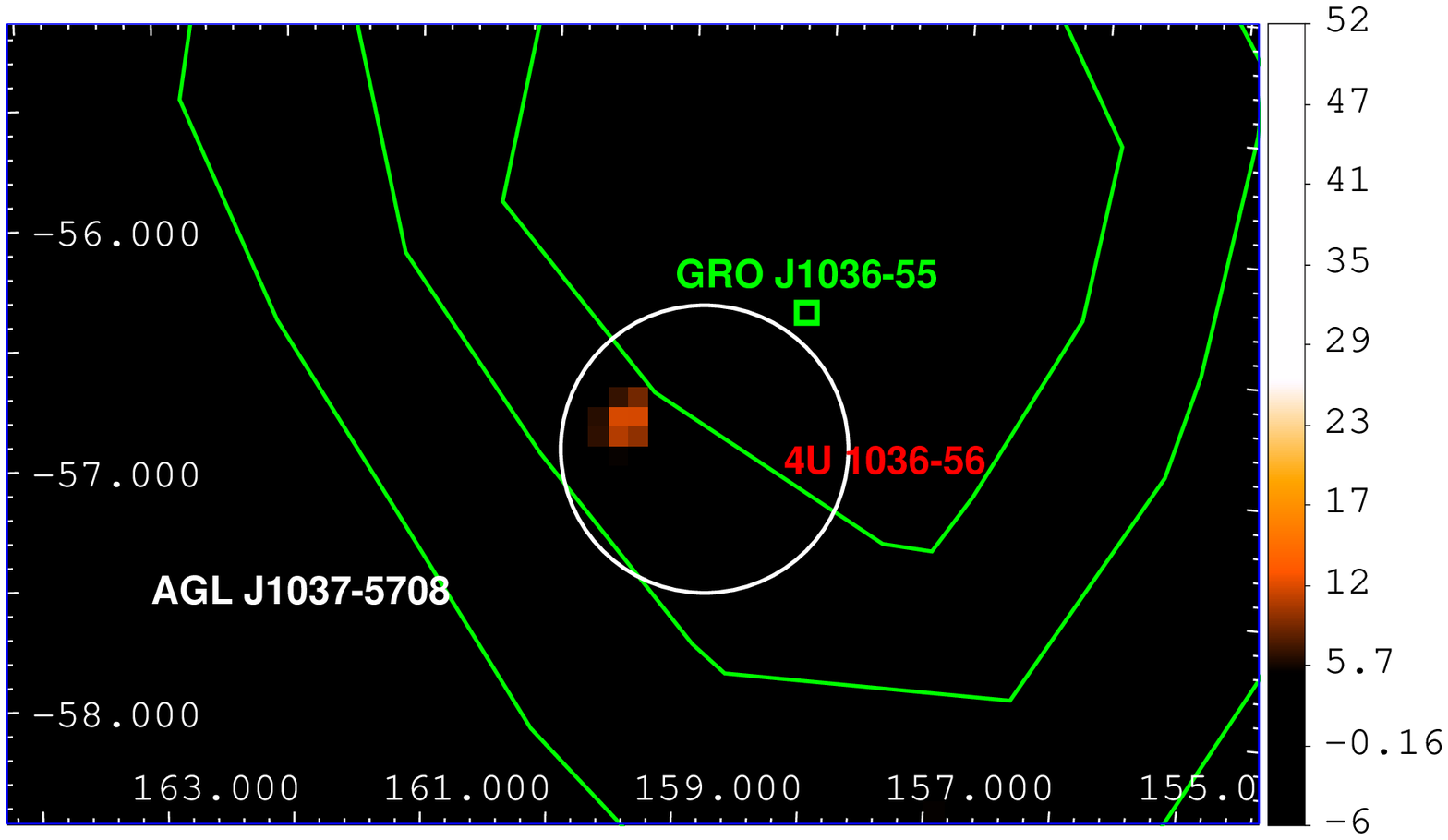}
\includegraphics[angle=0, scale=0.43] {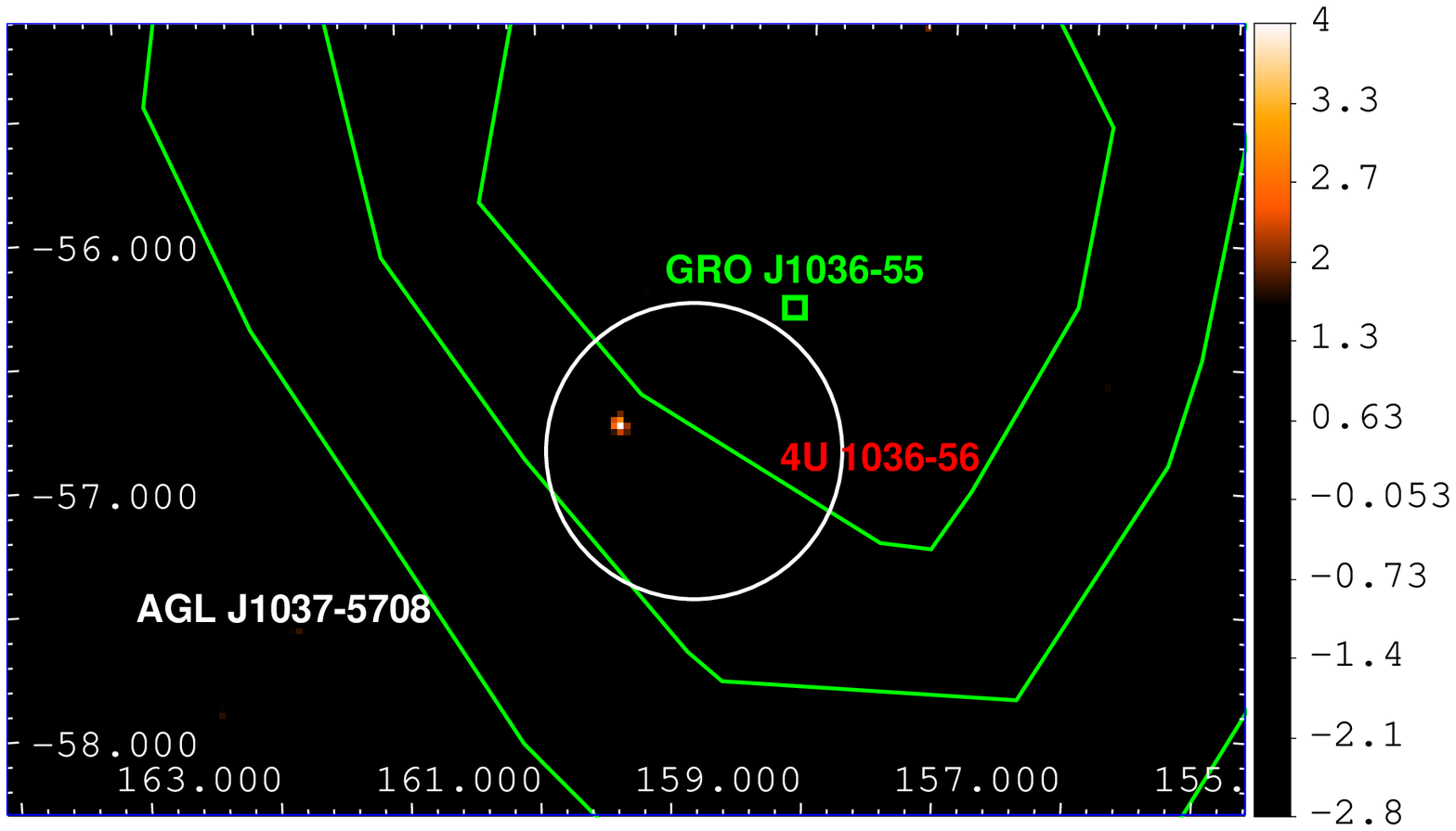}
\caption{Mosaic image of the \uu\ sky region, derived by
combining all \is data (18--60 keV, left panel) and
combining all \je data (3--20 keV, right panel). The
position of the transient {\it AGILE} source AGL J1037--5708 is plotted with its 95$\%$ error region (white). Another transient source, GRO J1036--55 is shown in sky region with its 1, 2, 3 $\sigma$ uncertainty location (green). The significance level is given by the color scale.  Corresponding significance and
color can be found in the right color bar. The X-- and Y--axis are RA and Dec. in
units of degrees.}
\label{mos-1}
\end{figure*}

\begin{figure*}[t]
\centering
\includegraphics[angle=0, scale=0.7] {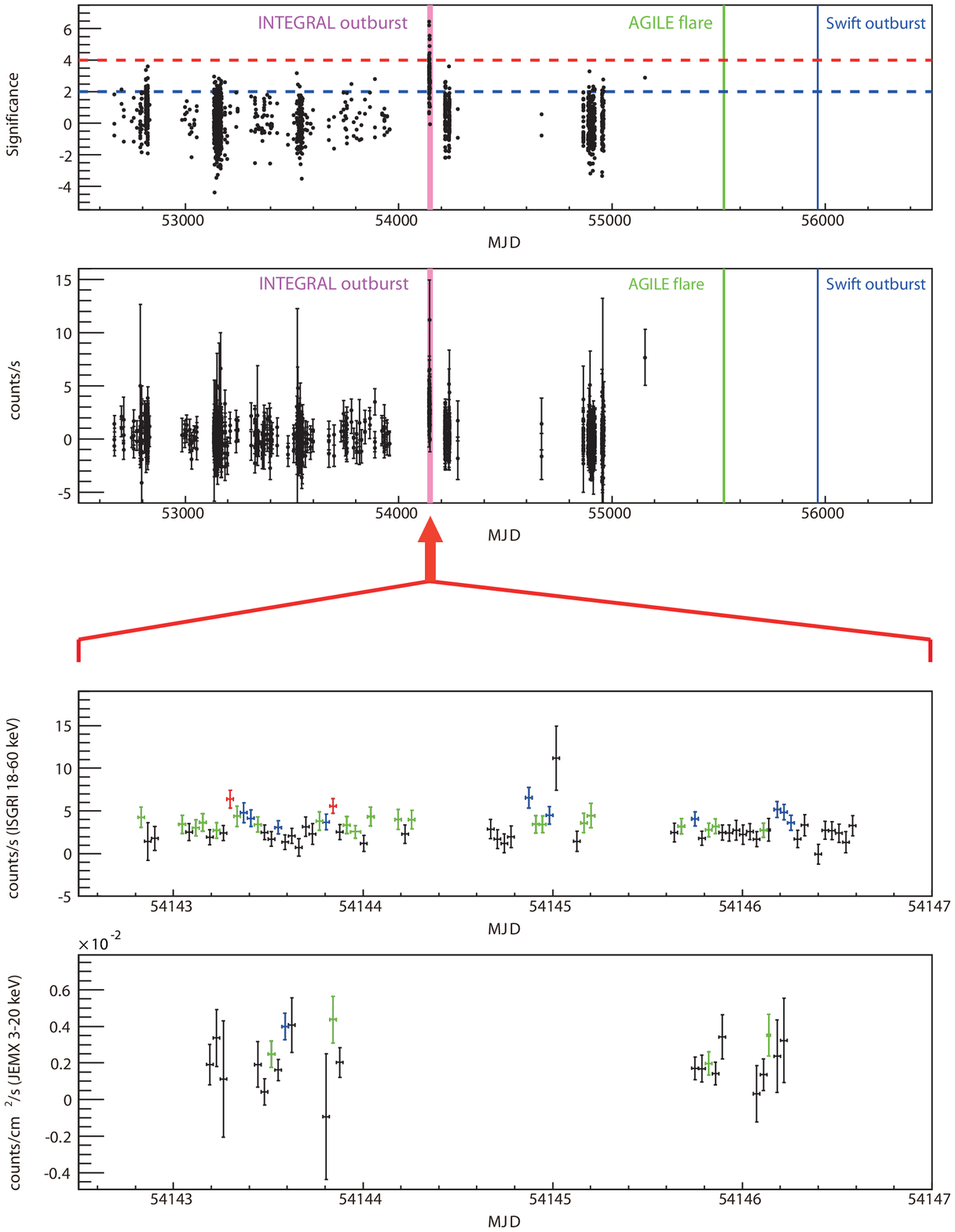}
\caption{Upper panel: Long--term significance (upper panel) and lightcurve (lower panel) of \uu on ScW timescales as seen by \is in the 18--60 keV band. The highlighted periods are the {\it INTEGRAL} outburst (pink), {\it AGILE} flare (green) and {\it Swift} outburst (blue). The dotted blue and red lines in the upper panel stand for the 2 and 4 $\sigma$ significance level. Lower panel: Zoomed \is (upper) and \je (lower) lightcurves on time scale of ScW. Red, blue, green and black points stand for ScWs above 5 $\sigma$, between 4--5 $\sigma$, between 3--4 $\sigma$ and below 3 $\sigma$, respectively.}
\label{mos-1}
\end{figure*}

\subsection{{\it INTEGRAL} outburst and hard-X-ray quiescence}

An outburst at MJD 54144 is significantly detected by \is having a significance of 30.4 $\sigma$ and an average intensity of 2.589$\pm$0.085 counts s$^{-1}$ in the 18--60 keV band, over a 199 ks exposure ($\sim$1/22 of the total exposure on the source).
The \je detection in the outburst period is significantly made at 10.1 $\sigma$ with an average intensity of $0.194\pm0.019\times10^{-2}$ counts/{cm$^{2}$}/s in the 3--20 keV band.
Out of the outburst period,  \uu is detected by \is only with a significance of 5.7 $\sigma$ under a total exposure time of 4.4 Ms in the 18--60 keV band.
The average flux is 0.094$\pm$0.018 counts s$^{-1}$ in the 18--60 keV band (27 times dimmer than in outburst).
\je does not detect \uu during quiescence, yielding only 2.58 $\sigma$ in the 3-20 keV, under a total exposure of 0.99 Ms. The mosaic images of the outburst and the quiescent period is show in Figure 3.

\begin{figure*}[t]
\centering
\includegraphics[angle=0, scale=0.43] {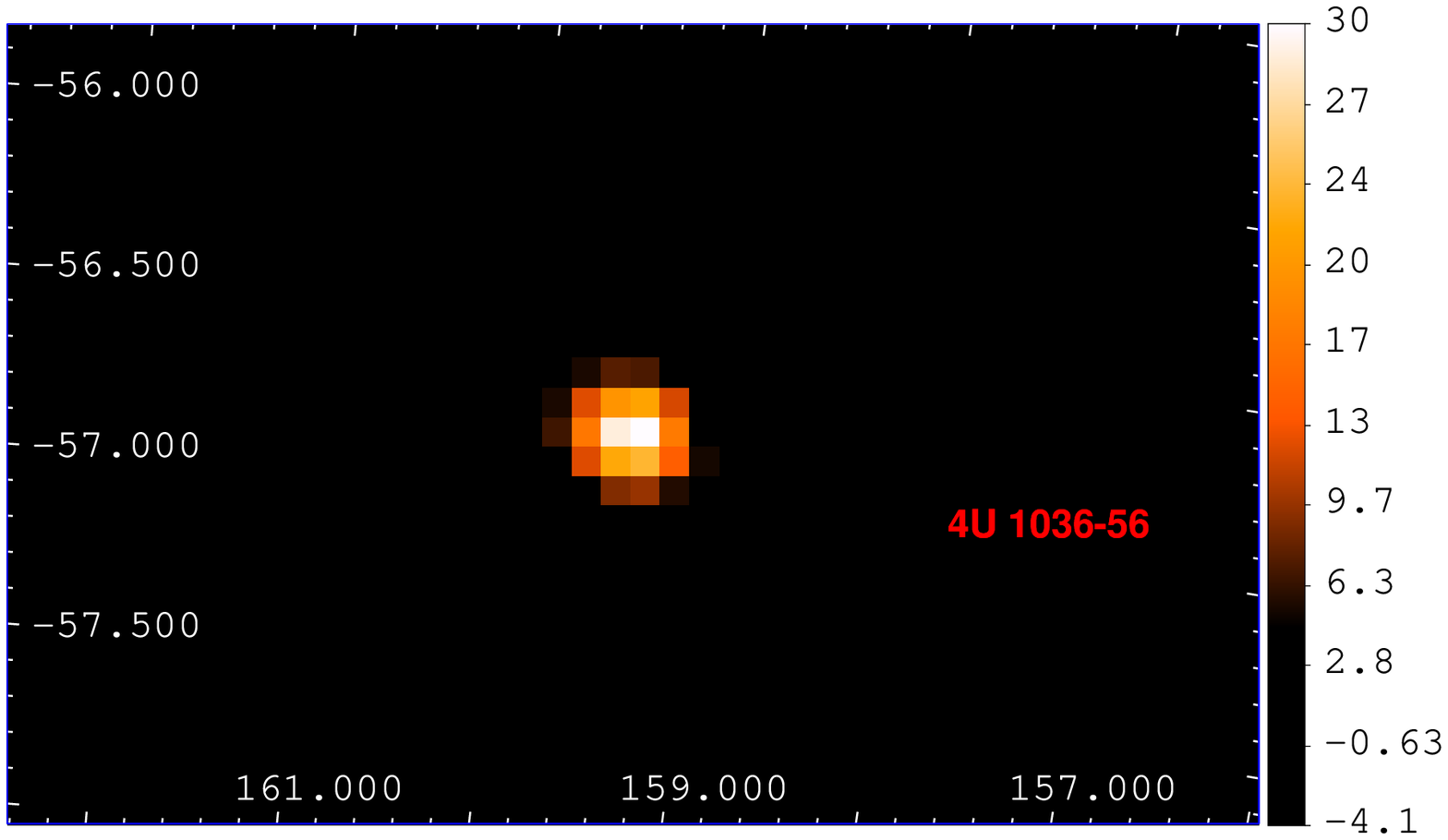}
\includegraphics[angle=0, scale=0.43] {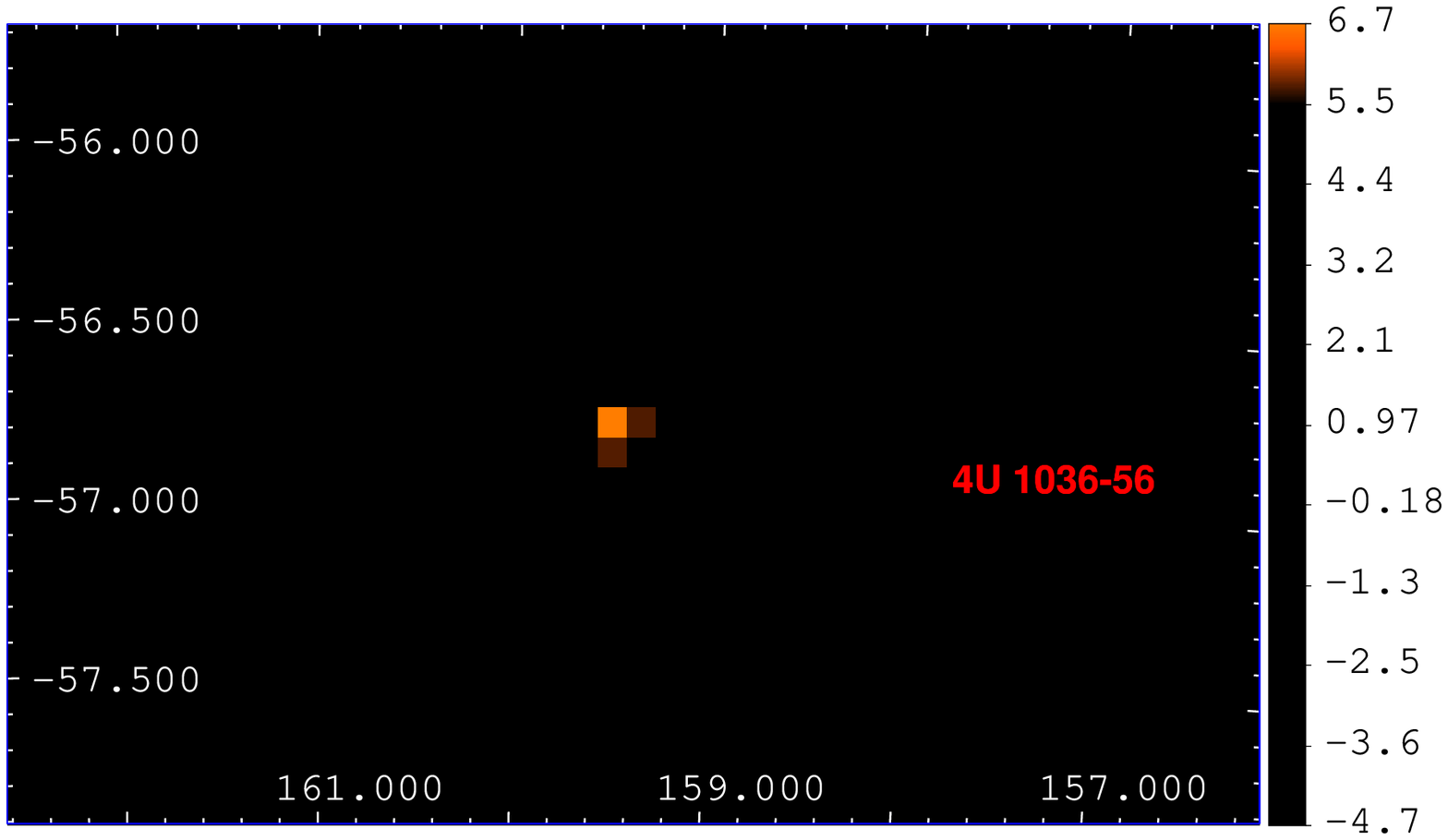}
\caption{The \is mosaic image in 18-60 keV of \uu\ during {\it INTEGRAL} outburst (left) and quiescence (right) periods. The significance level is given by the color scale. Corresponding significance and color can be found in the right color bar. The X-- and Y--axis are RA and Dec. in units of degrees.}
\label{mos-1}
\end{figure*}

In Figure 2 we show the zoomed lightcurve of the {\it INTEGRAL} outburst from \is and JEM--X. The counts rate are extracted from the position of the source in individual images on the ScW time--scale. Red, blue, green and black points stand for ScWs with significance above 5 $\sigma$, between 4--5 $\sigma$, between 3--4 $\sigma$ and below 3 $\sigma$, respectively. A constant fit to \je lightcurve yields an average flux of 0.204 $\pm$ 0.019 counts$\times10^{-2}$/cm$^{2}$/ s and a reduced $\chi^{2}$ of 1.3 {(21 D.O.F)}. Also, a constant fit to \is lightcurve along the outburst yields an average flux of 2.98 $\pm$0.12 counts s$^{-1}$ and a reduced $\chi^{2}$ of 1.5 (70 D.O.F). Both the \je and \is lightcurve does not hint for variability (less than 3 sigma), probably due to the large error bars.

Based on the significant detection of {\it INTEGRAL} outburst, we could extract an energy spectrum both from \je and \is and  perform a simultaneous spectral analysis (see Figure 4). We use an absorbed blackbody, an absorbed powerlaw, and an absorbed cutoff powerlaw model to fit the combined \je and \is energy spectrum. The hydrogen column density is fixed to {3.36} $\times$ $10^{22}$ cm$^{-2}$ following the value we derived from the {\it Swift} observation of the 2012 outburst (see next section for detail). We have found that only an absorbed cutoff powerlaw model could yield an acceptable fit. Its reduced $\chi^2 (dof)$ is 0.80 (8)
compared with 8.587 (9) and 3.555 (9) for absorbed blackbody and power law, respectively.
According to an F-test, the probability of refusing the cutoff to a simple power-law is 4.78$\times$ $10^{-4}$, corresponding to a significance of 3.5 $\sigma$.
The result of the fitting parameters are shown in Table 2. Based on the spectrum parameters of the {\it INTEGRAL} outburst and assuming a source distance of 5 kpc (Motch et al. 1997), the luminosity derived in 2--10 keV band is ${5.16}^{+0.72}_{-0.63}\times$ $10^{35}$ erg s$^{-1}$.

\begin{table*}
\centering
\scriptsize
\caption{ Summary of fitting parameters for the {\it INTEGRAL} outburst, the quiescent period, and the
{\it Swift} outburst. PL (BB) stands for power-law (blackbody) models, respectively.}
\begin{tabular}{lllllll}
\\
\hline
\hline
\\
Period & Hydrogen      & Photon   & Cutoff       & Blackbody         & Unabsorbed Flux & Reduced $\chi^{2}$ (D.O.F) \\
        &column density &  Index      &    Energy &     temperature     & ($10^{-10}$ erg cm$^{-2}$ s$^{-1}$)  &            \\
        &$10^{22}$ cm$^{-2}$  &        &   (keV)      &   (keV)   & &                                                   \\
\\
\hline
\\
{\it INTEGRAL} &\\
      &                &          &               &                &(18--60 keV) & \\
\\
outburst &3.36    (fixed)   &  1.12$^{+0.25}_{-0.25}$ & 26.49$^{+7.2}_{-4.8}$ & --- &2.09$^{+0.40}_{-0.35}$ & 0.80 (8)\\
\\
quiescence & ---  & 3.14$^{+0.63}_{-0.53}$&--- &--- &0.058$^{+0.009}_{-0.009}$ &0.03 (1)\\
\\
\hline
\\
{\it Swift} &\\
      &                &          &               &                &(2--10 keV)&\\
\\
PL  &3.36$^{+0.72}_{-0.53}$   & 1.08$^{+0.23}_{-0.18}$ & --- & ---                       &  0.868$^{+0.047}_{-0.042}$ & 0.699 (34)\\
\\
BB  &1.41$^{+0.41}_{-0.37}$   & ---                      & --- & 1.85$^{+0.16}_{-0.14}$  &  0.698$^{+0.039}_{-0.038}$ & 0.617 (34)\\

\\
\hline
\\
\end{tabular}
\label{expo-hard2}
\end{table*}

As a result of the low significance of \je detection corresponding to the
out-of-outburst period, we could not extract a meaningful energy spectrum or lightcurve.  We extracted the spectrum from \is in large energy bins directly from mosaic images.
The counts rate is low but due to the long exposure time, there are enough counts accumulated to allow for a $\chi^{2}$ statistics fitting.
Because of the limited energy bins and the low statistics, a simple power law is fitted to the spectrum (Figure 4, lower blue line). Details are given in Table {2}.
 Since there is only 1 degree of freedom in the spectral fitting, the fitting is not good and results are only indicative.
Despite the obvious caveats of the spectrum determination,
the derived flux of \uu when in quiescence is $\sim 36$ times lower than in the outburst. This is consistent with the dynamic range of $\sim 30$ derived from count rate (18--60 keV). Since \uu is quite weak during the quiescent period, this is very likely the reason why the source was not reported in the second IBIS catalogue (Bird et al. 2006). On the contrary, the source was listed in the subsequent third and fourth IBIS catalogues (Bird et al. 2007, 2010) with a similar 18--60 keV average flux.

\subsection{{\it Swift} outburst}

On MJD 55960 (2012 Feb. 3) an outburst from \uu was caught by {\it Swift}/BAT (Krimm et al 2012). \uu was detected with a rate of 0.0040 $\pm$ 0.0008 counts s$^{-1}$ cm$^{-2}$ ($\sim$20 mCrab) in the 15--50 keV at the early state of the out burst. The peak was reached on 2012 Feb. 6 with a flux of 0.0067$\pm$ 0.0019 counts s$^{-1}$ cm$^{-2}$ ($\sim$30 mCrab). The detection of \uu went on until Feb. 13 (MJD 55970), but afterwards \uu went undetected (1 $\sigma$ upper limit of 0.001 counts sec$^{-1} $cm$^{-2}$). This outburst lasted for about 10 days. The {\it Swift}/XRT observation was performed on 2012 Feb. 17. We have carried out spectral and timing analysis of {\it Swift} outburst. The spectrum in 0.3--10 keV could be well fitted by an absorbed powerlaw or an absorbed blackbody. Fitting parameters are shown in Table {2} {and are consistent with the spectral results of Krimm et al. (2012).}
Also, assuming a source distance of 5 kpc (Motch et al. 1997), the luminosity derived for the {\it Swift} outburst in the 2--10 keV band is about {2.60}$\times 10^{35}$ erg s$^{-1}$.

We have also carried out timing analysis of {\it Swift} outburst. A 50 s-binned lightcurve of \uu in the 0.3-10 keV band is shown in Figure {5}. The flux shows variability that matches the pulse period. To search for a periodic signal in the lightcurve data, we generated the power spectrum. A significant signal is detected at 854.75$\pm$4.39 s (1 $\sigma$ errors), beyond the white and the red noise (Figure 5, middle panel; since the
red--noise is below the white--noise component, only the white--noise is shown). The derived pulse period is consistent with the 854.3$\pm0.2$ seconds period determined by the \textit{XMM-Newton} observation (La Palombara et al. 2009). The relatively large error of pulse period and a wide peak seen in the power spectrum may be due to the limited pulse cycles (only about 4) covered in lightcurve, which lead to uncertainties. We fold the lightcurve at the 854.75 seconds pulse period and derived the pulse profile (Figure {5}, lower panel). Phase zero is arbitrarily taken as the starting time of lightcurve. The pulsation exhibits a single peak profile, similar with the pulse profile in Reig \& Roche (1999) and La Palombara et al (2009). Fitting a
sinusoidal function to the pulse profile yielding a pulse fraction (ratio of sinusoidal amplitude to mean count rate) of 36.2$\pm$6.7$\%$ (Figure {5}, lower panel).

\begin{figure*}[t]
\centering
\includegraphics[angle=270, scale=0.6] {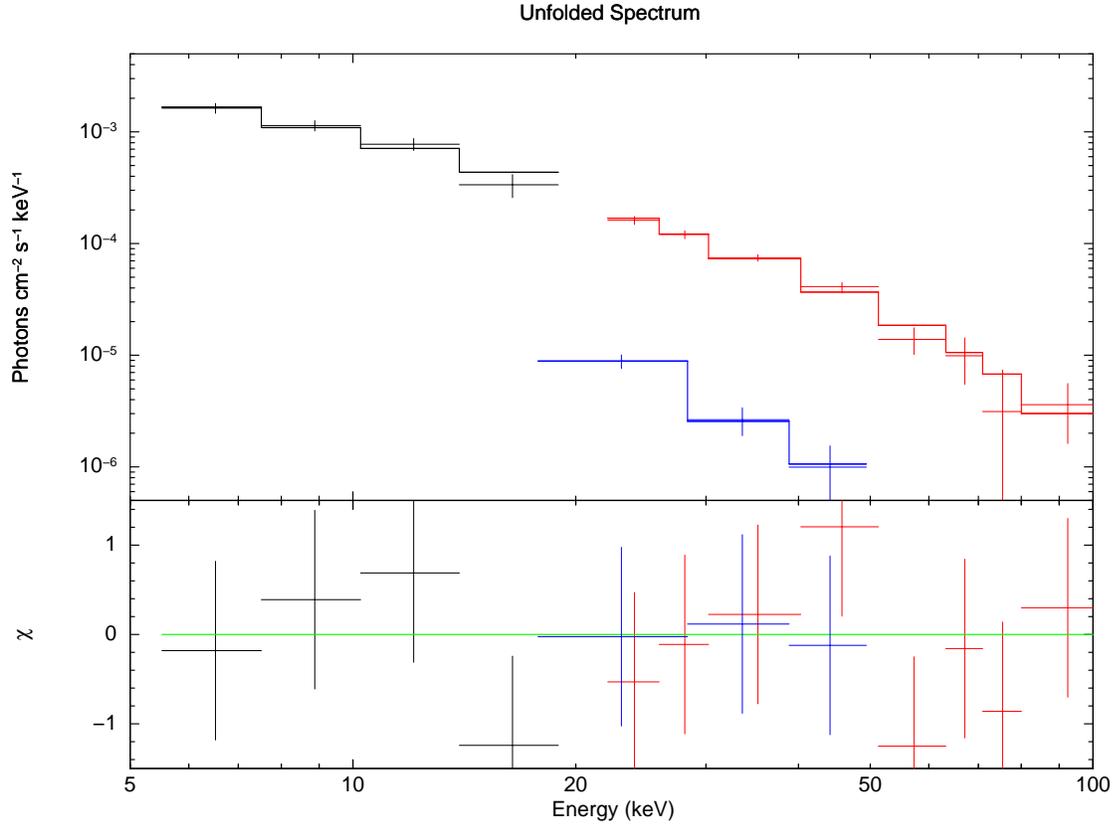}
\caption{Upper: unfolded {\it INTEGRAL} energy spectrum of the outburst (upper line) and quiescence period (lower line). The black points are from \je while red ones are from IBIS/ISGRI during outburst. The blue points are from IBIS/ISGRI during quiescence period. Lower: residual of the fit. Black and red points are from \je and IBIS/ISGRI during outburst while blue points are from IBIS/ISGRI during quiescence period.}
\label{mos-1}
\end{figure*}

\begin{figure}[t]
\centering

\includegraphics[angle=270, scale=0.31] {pileup_4pixel_bary_2.51s_0.3-10kev_corr_net.ps}

\includegraphics[angle=0, scale=0.41] {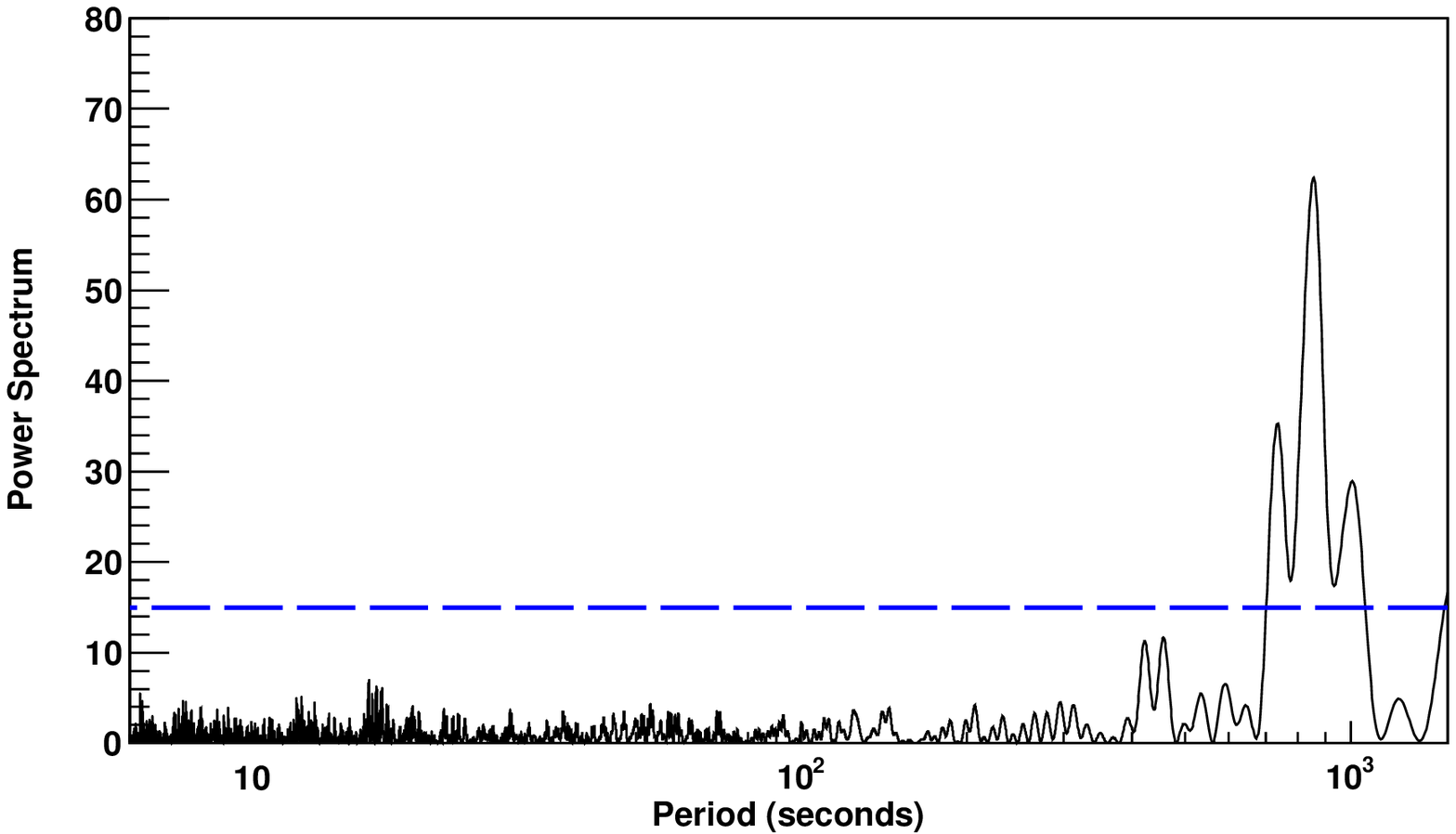}

\includegraphics[angle=0, scale=0.41] {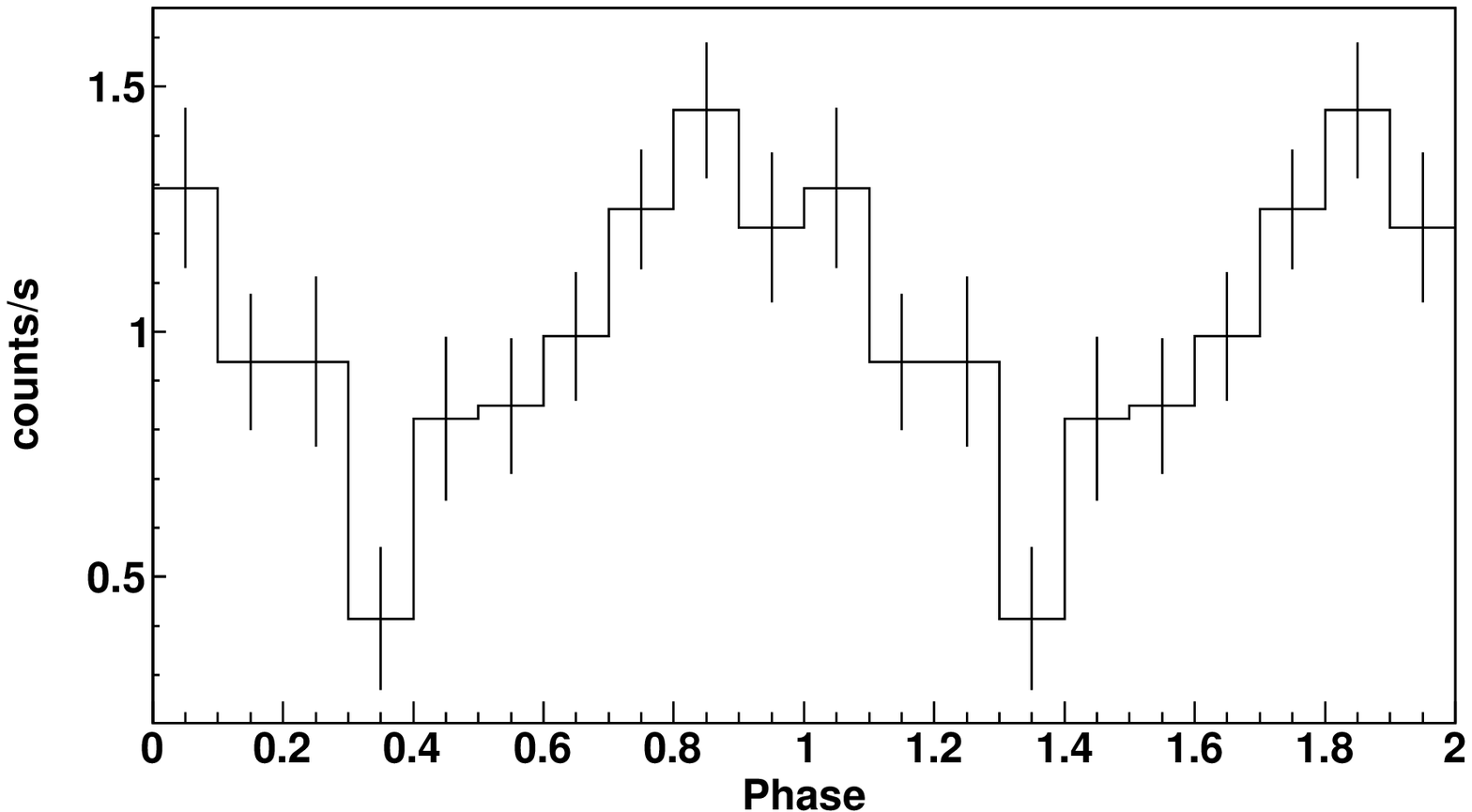}

\caption{Upper: {\it Swift}/XRT lightcurve of \uu (0.3--10 keV, 50 s bins) during the observation on 2012 Feb. 17. Middle: power spectrum (for a 2.5 s bins binned lightcurve in 0.3--10 keV) and 99.99$\%$ white noise level (in blue). Lower: profile folded at 854.75 s period. Phase zero is the starting time of the lightcurve.
}
\label{mos-1}
\end{figure}

\section{Discussion}

In this paper we have reported on {\it INTEGRAL} and {\it Swift} observations of 4U 1036--56.
We have found that
 \uu is a weak hard X-ray source spending most of the time in a quiescent state with an average flux of 5.8 $\times10^{-12}$ erg  cm$^{-2}$ s$^{-1}$ (18--60 keV). Occasionally, we have seen, \uu may exhibit  outbursts which last for several days. The flux in the outburst we analyzed reached 2.09 $\times10^{-10}$ erg cm$^{-2}$ s$^{-1}$ (18-60 keV) band, implying a dynamical range of $\sim 36$.

\uu is the only hard X-ray source in the uncertainty contour of the $\gamma$-ray transients AGL J1037-5708,
as well as on the admittedly worse-localized GRO J1036--55.
AGL J1037-5708 was discovered on 2010--11--27 during a flare lasting for 3 days (Bulgarelli et al. 2010). The positional coincidence as well as the variability timescales make it possible to entertain the hypothesis that the high-energy transients and \uu are related.
The possible association between a binary and an {\it AGILE} transient, under the weight of similar, exploratory arguments has been mentioned before (see Table 1 and references therein). Some of the suggested counterparts for the low number of transients found are SFXTs.
These are characterized by faster X-ray flares ($\sim$ hours) and larger dynamical ranges ($10^{3}$) than the ones found for 4U 1036--56.

Several authors proposed that it is theoretically feasible that HMXBs produce  $\gamma$-ray emission during  periods of X-ray activity induced by accretion (see, e.g., Bednarek 2009 for a leptonic and {Romero et al. 2001} and Orellana et al. 2007 for a hadronic model). The latter has been put to the test
by observations of the Be/pulsar binary 1A 0535+262 during a giant X-ray outburst (Acciari et al. 2011).
Whereas it is beyond the scope of this paper to produce a detailed theoretical model of the source, we consider next whether it is in principle plausible that \uu and the {\it AGILE} flares are related. We focus here on the consideration of the leptonic model.

The idea  is similar to that used for propellers or $\gamma$-ray binaries which might be in a flip-flop between a propeller and an ejector state (see, e.g., {Bednarek 2009, Bednarek \& Pabich 2011, Torres et al. 2012)}.
The dense wind of a massive star can be partially captured
by a neutron star inside a compact binary system. If the neutron star is rotating
slowly, as is the case of 4U 1036--56, the matter from the stellar wind can penetrate the inner
neutron star magnetosphere. This matter can be directed towards the neutron surface following magnetic lines.
At some distance from the neutron star, a turbulent and magnetized transition region (a distance
known as the Alfven or magnetic radius)
is formed due to the balance between the magnetic pressure and the pressure of
accreting matter. This region, the position of which can be computed as (see Bednarek 2009 and references therein)
\be
R_A \sim 4 \times 10^8 B_{12}^{4/7} \dot M_{16}^{-2/7} {\rm cm},
\ee
where
$B_{12}$  is the magnetic field at the neutron-star surface in units of  $10^{12}$ G and $\dot M_{16}$ is the accreted mass
in units of $10^{16}$ g s$^{-1}$, may
provide good conditions for acceleration of particles to relativistic energies. Note that the fiducial values entirely  depend on the properties of the neutron star and the orbit of the binary can easily be significantly off the chosen scales.

\begin{figure}[t!]
\centering
\includegraphics[angle=0, scale=0.48] {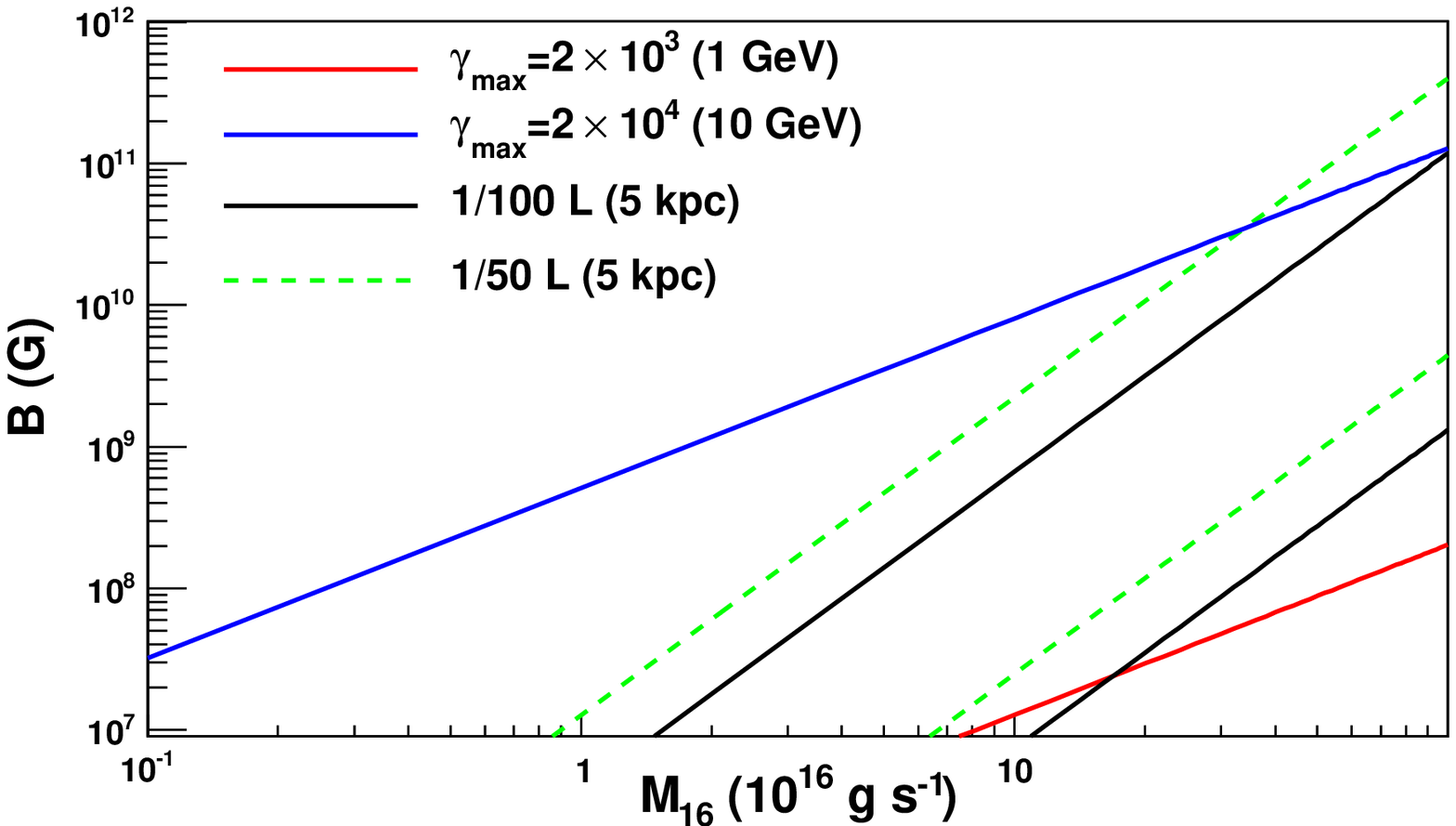}
\includegraphics[angle=0, scale=0.48] {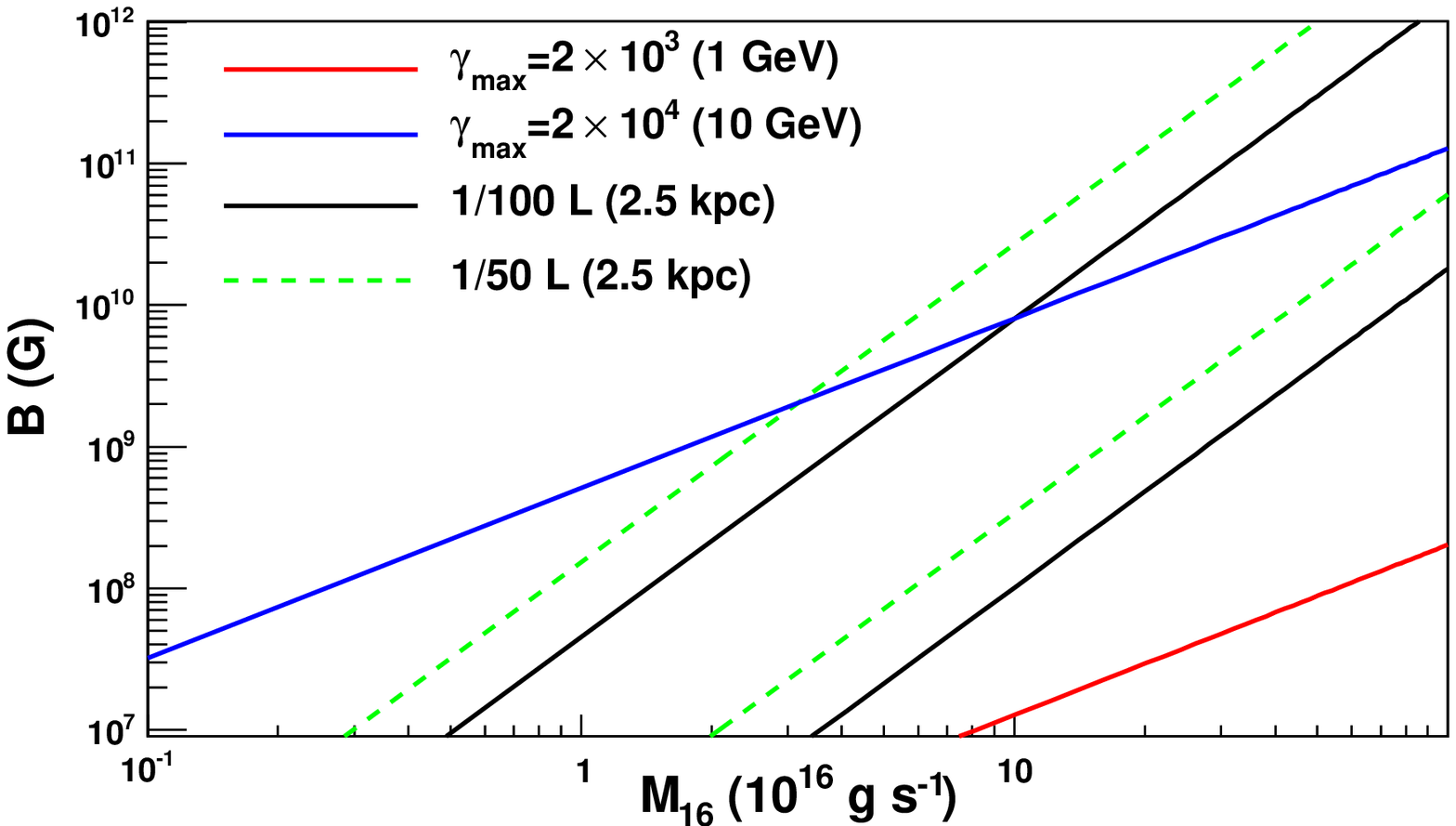}
\caption{Phase space for the magnetic field of the neutron star in \uu and the
accretion rate at the magnetic radius for which electrons of 1 and 10
GeV can be accelerated with sufficient power sustain the AGILE flare. The upper panel is for
a distance estimation of 5 kpc while the lower panel is for a distance estimation of 2.5 kpc.
Black and green lines stand for the observed AGILE MeV emission in the phase space, assuming it is
$1/100$ or $1/50$ of the accretion power at Alfven radius ($R_A$), respectively. The space
constrained by two parallel lines for each color (green $\&$ black) represents the observed
AGILE MeV emission as a result of a possible power-law index ranging from 2.5 to 4.0. The red and blue
lines stand that electrons could be accelerated to 1 and 10 GeV respectively through
the mechanism proposed in this paper, or even beyond these energies above the corresponding
lines. The intersection area of above these constraints stand for the observed AGILE MeV
emission which is possibly originated from the mechanism we proposed. See text for more details}
\label{cons}
\end{figure}

For accretion to occur, three conditions must be satisfied: a) the magnetic radius $R_A$ is inside the light cylinder $R_{lc}=cP/2\pi$, where $P$ is the neutron star period, and which imposes a lower limit  on $P$; b) the rotational velocity of the magnetosphere at $R_A$ is smaller than the Keplerian velocity of the accreting matter (so no propeller is allowed);
and c) the magnetic radius is also smaller than the capture radius of the matter of the stellar wind. All three conditions are met for the known parameters
of 4U 1036--56. In particular, the large pulse period found secures the neutron star is in the accreting state.
By equating the electron acceleration timescales with that of the losses, in particular, synchrotron losses (albeit electrons with energies lower than GeV could also be affected by significant inverse Compton losses) for the higher energy particles,
one obtains the maximum Lorentz factor of the electrons as
\be
\gamma_{max} \sim 3 \times 10^5 \zeta_{-1}^{1/2} B_{12}^{5/14} \dot M_{16}^{-3/7},
\label{gm}
\ee
where $\zeta_{-1}$ is the acceleration efficiency, a dimensionless number in units of 0.1 (which value is unknown, e.g., Aharonian et al. 2002 considered values of $10^{-2}-10^{-4}$, Bednarek 2009 and others, used the prior
scale).
These electrons can emit synchrotron photons of characteristic energy
$\epsilon_s = m_e c^2 (B_A/B_{cr}) \gamma_{max}^2 \sim
1.9  \zeta_{-1} {\rm MeV}.
$
 Apart from the usual magnitudes of the speed of light and the mass of the electron, $B_A$ is the magnetic field at the position of $R_A$, obtained through the dipolar formula, and $B_{cr}$ is the critical magnetic field $4.4 \times 10^{13}$ G. Relatively weak X-ray sources (low accretion) can accelerate particles to higher energies than
the more powerful X-ray binaries. For the latter, tens of GeV can be considered as a safe upper limit for the maximum energy of the accelerated electrons. The precise value of maximal energies, we emphasize, depend on the several parameters describing the neutron star and the orbit. All in all, synchrotron emission can extend to the hard X-ray regime and beyond.


In a propeller case, the maximum power available for the acceleration of electrons is limited by the energy extracted from the rotating neutron star by the infalling matter, which can be estimated as
(Bednarek 2009) as $L_{rot} \sim M_{acc} v_{rot}^2 / 2 \sim 3 \times 10^{34} B_{12}^{8/7} \dot M_{16}^{3/7} P_{1}^{-2}$ erg s$^{-1}$, with $v_{rot}$ the rotational velocity of the magnetosphere. If this would also be the case for accreting systems,
the long period of \uu (and essentially any other neutron star in a high mass accreting binary, for which $P>100$ s are common) would imply too low a power to sustain
typical flare fluxes. However, in accreting scenarios,
given the rotational velocity of the magnetosphere is slower than the Keplerian velocity of the accreting matter,
the gravitational energy and the specific angular momentum of the infalling matter flow from the latter to the neutron star, and not viceversa.
This is consistent with the observation of pulsations and with the fact that the neutron star is indeed spun up. One can then consider that the maximum power available for the acceleration of electrons
is a fraction of the accretion power at $R_A$; i.e.
\be
L = GM \dot M / R_A \sim 4.6 \times 10^{33} B_{12}^{-4/7} \dot M_{16}^{9/7}.
\label{pow}
\ee
If this is the case, in a low accretion rate mode, the source barely has enough power to appear as a source in MeV observations; it shows up only as X-ray / hard X-ray source as a result of the accretion process (i.e., when matter actually falls in the neutron star surface).
On the contrary, hard X-ray or higher energies flares
may be observable only when the accretion rate increases enough (e.g., by an increase in the mass loss rate, or accretion of a wind clump) so that the fraction of the power that can be converted into relativistic electrons at $R_A$ may lead to significant non-thermal luminosity.
However, if the accretion rate increases enough, the maximum electron energies decreases, as shown above by Eq. (\ref{gm}), actually reducing the phase space in which, for instance, an {\it AGILE} source may be detected.

Due to the {\it AGILE} energy resolution, the $E > 100$ MeV flux
contains a large contribution from sub-100 MeV photons (see, e.g., { Longo  2010)}.
Actually, given that {\it Fermi}-LAT (optimized at 1 GeV) did not detect this
and other {\it AGILE} transients, it is reasonable to suppose that these flares are sub--100 MeV ones.
Nevertheless,
the photon flux of the AGL J1037-5708 flare is quite intense (Bulgarelli et al. 2010).
To give some examples, assuming it corresponds to an energy range from 30 to 200 MeV, and that its spectrum (which is not known)
is very steep (corresponding to the fact that {\it Fermi}-LAT has not detected it)
one obtains the range
$ (7.6 \times 10^{35}  - 5.8 \times 10^{34}) (D / 5 \, {\rm kpc})^2$ erg s$^{-1}$, for a power-law photon spectral index of 2.5 and 4.0, respectively. This power should only be a fraction of Eq. (\ref{pow}); which also imposes constraints on the magnetic field and accretion rate. An example of these constraints are plotted in Figure \ref{cons}, upper panel. The red and blue lines correspond to defining a $\gamma_{max}$ equal to 1 and 10 GeV respectively; i.e., above the corresponding lines, the electrons can be accelerated
beyond those energies.
The parallel sets of black and green lines correspond to the constraints imposed by asking that the {\it AGILE} flare luminosity interval (as given above, depending on the spectrum) is 1/100 and 1/50 of the total power given in Eq. (\ref{pow}).
Pending the development of detailed models, including the computation of opacities,
it then seems that it is a priori possible (although not necessarily preferred)
that \uu and the $\gamma$-ray transients could be related;
particularly if the magnetic field is low.
However, for a typical magnetic field of 10$^{12}$ G the accreted mass seems prohibitively high in order to sustain the emission by the mechanism described.
We should also take into account that the distance to \uu is quoted as 5 kpc but it can sustain an uncertainty
of 25 to 50\% (Motch et al. 1997). A distance of 2.5 kpc, for instance,
would translate into a change of  $\gamma$-ray luminosity
between
($1.7 \times 10^{35}  - 1.4 \times 10^{34}$) erg s$^{-1}$
for the same range of spectral slopes. This would make the possibility of an association between \uu and AGL J1037--5708
more plausible. This is shown in the lower panel of Figure \ref{cons}.

\acknowledgements

This work was supported by the grants AYA2009-07391 and SGR2009-811, as well as the Formosa program TW2010005 and iLINK program 2011-0303. We acknowledge support from the National Natural Science Foundation of China,
the CAS key Project KJCX2-YW-T03, 973 program 2009CB824800 and NSFC-11103020,
11133002, 10725313, 10521001, 10733010, 10821061, 11073021, 11173023.
We thank A. Camero-Arranz and W. Bednarek for comments.

\end{document}